\documentclass[preprint,aps,prd,superscriptaddress,showpacs,showkeys,nofootinbib,notitlepage ]{revtex4-1}
\usepackage{graphicx}
\usepackage{amsmath}
\usepackage{longtable}
\usepackage{color}

\newcommand{\be}{\begin{equation}}
\newcommand{\ee}{\end{equation}}
\newcommand{\bea}{\begin{eqnarray}}
\newcommand{\eea}{\end{eqnarray}}
\newcommand{\etal}{et al.}

\begin{document}

\bibliographystyle{apsrev}

\title{Equivalence of the Fermat potential and the lensing potential approaches to computing the integrated Sachs-Wolfe effect}

\author{R. Kantowski}
\email{kantowski@ou.edu}
\affiliation{Homer L.~Dodge Department~of  Physics and Astronomy, University of
Oklahoma, 440 West Brooks,  Norman, OK 73019, USA}

\author{B. Chen}
\email{bchen3@fsu.edu}
\affiliation{Research Computing Center, 
Florida State University, Tallahassee, FL 32306, USA}
\author{X. Dai}
\email{xdai@ou.edu}
\affiliation{Homer L.~Dodge Department~of  Physics and Astronomy, University of
Oklahoma, 440 West Brooks,   Norman, OK 73019, USA}
\date{\today}
\begin{abstract}
We show in detail that the recently derived expression  for evaluating the integrated Sachs-Wolfe (ISW) temperature shift in the cosmic microwave background  (CMB) caused by individual embedded (compensated) lenses is equivalent to the conventional approach for flat background cosmologies.
The conventional approach requires evaluating an integral of the time derivative of the lensing potential, whereas the new Fermat potential approach is simpler and only requires taking a  derivative of the potential part of the time delay.
\end{abstract}

\pacs{98.62.Sb, 98.65.Dx, 98.80.-k}

\keywords{General Relativity; Cosmology; Gravitational Lensing;}

\maketitle

\section{Introduction}
 A renewed interest in the late time integrated Sachs-Wolfe (ISW) effect \cite{Sachs67}, also known as the Rees-Sciama (RS) effect \cite{Rees68}, has recently arisen  because hot and cold spots in the CMB temperature have been associated with some known large scale structures---galaxy clusters and cosmic voids \cite{Granett08,Granett08b,Planck14}. The ISW/RS effect is the shifting of the CMB temperature when viewed through one or more gravitational lenses. In this paper we are interested in the ISW/RS effect caused by a single embedded lens. The actual shift depends on details of the lens profile and its kinematics where probed by the transiting CMB photons.
By modeling cluster and void density profiles and internal motions, and by adjusting cluster masses and void depths, observed temperature excesses/deficits can be matched by ISW predictions \cite{Inoue06,Rudnick07,Nadathur12,Hernandez10,Ilic13,Cai14,Chen15b,Chen15c}.
Several proposals exist to use  lensing of the CMB to determine properties of these clusters  and voids as well as the cosmological parameters \cite{Lavaux12,Melin14,Chantavat14,Hamaus14,Kantowski15}.
While recently developing the embedded lens theory, which could also be called the Swiss cheese lens theory, or at lowest order, the compensated lens theory \cite{Kantowski10,Chen10,Chen11,Kantowski12,Kantowski13}, we discovered a relatively simple expression giving the ISW temperature shift as a derivative of the ``Fermat potential" of the lens \cite{Chen15a}.
The conventional approach to determine the ISW effect is somewhat more complicated and requires integration of the time derivative  of the ``lensing potential" along the transiting CMB photon's path \cite{Sachs67, Rees68}.
Our method of evaluating the ISW effect is  directly related to the projected lens' mass profile and hence more transparent than the conventional approach.
It is simpler to use and requires the construction of only one single function, the potential part of the time delay \cite{Cooke75}.
What we show in this paper is that the two methods actually give the same results when applied to the same compensated lens embedded in a spatially flat $\Lambda$CDM cosmology. Proof of equivalence for non-flat backgrounds is more complicated and not carried out here.
In Secs.\,\ref{sec:Embedded} and \ref{sec:Fermat} we describe the structure of an embedded lens and the associated Fermat potential.  We next review the expression giving the ISW effect on the CMB temperature Eq.\,(\ref{calT}) as a redshift derivative of the Fermat potential in Sec.\,\ref{sec:ISW-Fermat} and Eq.\,(\ref{calTconv}) as an integral of the lensing potential in  Sec.\,\ref{sec:ISW-Potential}.
In Secs.\,\ref{sec:Equivalence} we show that these two expressions give exactly the same results for flat $\Lambda$CDM cosmology. 
In the Appendix we give  tables of various lens mass densities, their projected mass fractions, and their time delay $T_p$ functions and demonstrate how simple it is to make linear superpositions of lenses masses.

\section{Embedded lenses}\label{sec:Embedded}
The logic for  using embedded lenses is simple, by computing the mean density inside larger and  larger spheres centered on a density perturbation, a radius will be reached beyond which the mean density coincides with the FLRW background. This is a reasonable assumption for a density perturbation that grow primarily by gravity from small fluctuations in the early universe. The  masses of such perturbations increase at the expense of the depleted surrounding mass density. Hierarchal clustering and merging clearly complicates this simple picture and extending use for this theory to perturbations that are not locally embedded will be discussed in future work. 

\begin{figure*}
\includegraphics[width=0.9\textwidth,height=0.18\textheight]{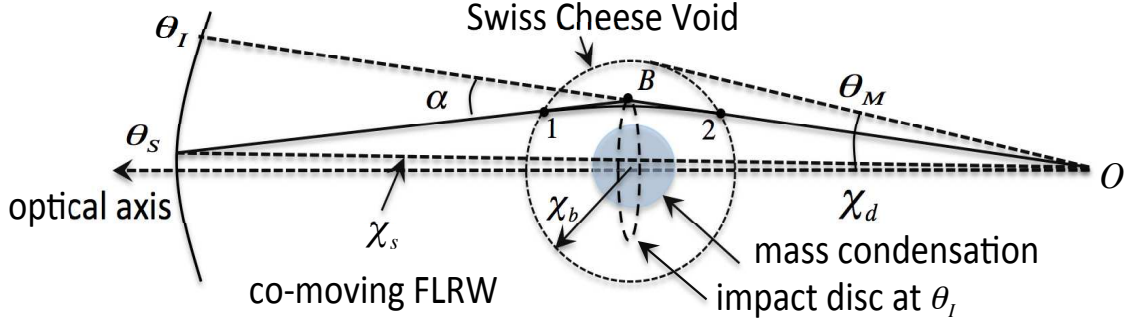}
\caption{ The comoving geometry of an embedded lens centered at redshift $1+z_d=R_0/R(t_d)$.
Angles $\theta_S$ and $\theta_I$ respectively, are  source and image angles;
${\chi}_d$ and ${\chi}_s$ are the comoving angular distances of the lens and the source from the observer.
The (constant) angular size of the void, in lowest order lensing theory,  is $\theta_M\equiv \chi_b/\chi_d$ where $\chi_b$ is the comoving radius of the Swiss cheese void.
	The physical radius of the deflecting lens depends on the central lensing time $t_d$ and is $r_d= R(t_d)\chi_b$.
	The shadowed area represents an embedded cluster.
	The dashed circle shows the impact disc of angular radius $\theta_I$, used to compute the included projected mass fraction $f(x)$ of the lens, see Eq.\,(\ref{T}).
         The equivalent figure for a void lens has a mass condensation surrounding a low density central region and a repulsive instead of attractive deflection angle $\alpha$.
}
\label{fig:cheese}
\end{figure*}

Because the embedded lens theory originated from the Swiss cheese models of general relativity (GR) \cite{Einstein45,Schucking54,Kantowski69} one can be confident of the correctness of its gravitational predictions.
An embedded lens  at redshift  $z_d$ is constructed by first removing a sphere from the background cosmology  whose boundary has a constant comoving angular radius $\chi_b$ producing a Swiss cheese void, see Fig.\,\ref{fig:cheese} and Eq.(\ref{metric}).
The removed mass $M$ is then replaced by an evolving spherically symmetric density in such a manner as to keep the Einstein equations satisfied inside and on the void's boundary to what ever accuracy is desired.
For example if the void has a physical radius $\chi_b R(t)$ at cosmic time $t$ and if the region just interior to the void boundary is a vacuum, GR requires that its geometry be described by the Kottler metric \cite{Kottler18} (Schwarzschild with a cosmological constant). The comoving radius of the outer void wall is related to the Schwarzschild radius of the embedded metric by
\be
r_s=\frac{2 GM}{c^2}=\frac{2 G}{c^2}\times\frac{4\pi}{3}\Bigl(\chi_bR(t)\Bigr)^3\overline{\rho}(t)=\Omega_m\left(\frac{H_0}{c}\right)^2\,\bigl(\chi_bR_0\bigr)^3.
\label{rs}
\ee
The current radius of the FLRW universe is taken to be $R_0=1$ by convention for spatially flat cosmologies and $\chi$ carries units of length.

The original Swiss cheese cosmologies filled part (or all) of the void's interior with the Schwarzschild metric and the remaining central part with a more dense homogeneous cosmology.  An interior FLRW cosmology bounded on the outside by the point mass gravity field has to satisfy a similar boundary condition to that of Eq.\,(\ref{rs}). Additional exact Swiss cheese models were constructed by filling the void with one of the Lema\^itre-Tolman-Bondi (LTB) models \cite{Lemaitre33,Tolman34,Bondi47}. In this paper we are only interested in the lowest order lensing properties and replace the void's mass by any spherical density perturbation whose net mass is $M$. Our only other constraint is that stresses and momentum densities within the lens are sufficiently small so as not to invalidate use of Newtonian perturbation  theory, see Eq.\,(\ref{metric}). Embedded models for physical voids must be surrounded by higher density regions and embedded cluster models surrounded by lower density regions.
Such linearized gravitational models are often referred to as compensated \cite{Nottale84,Martinez90,Panek92,Seljak96b,Sakai08,Valkenburg09}.

\section{The Fermat potential}\label{sec:Fermat}

For spherical density perturbations we have shown in \cite{Kantowski13,Chen15a} that to lowest order the gravitational lensing properties of an embedded lens can be completely described by its Fermat potential  (equivalent to the sum of the geometrical and potential time delays, $cT=c(T_g+T_p)$
 \bea
cT(\theta_S,\theta_I)&=& (1+z_d)\frac{D_dD_s}{D_{ds}}\Bigg[\frac{(\theta_S-\theta_I)^2}{2}
 +\theta_E^2\int_{x}^{1}\frac{f(z_d,x')-f_{\rm RW}(x')}{x'}dx'\Bigg].
\label{T}
\eea
Here $x\equiv\theta_I/\theta_M$ is the normalized image angle or equivalently the photon's fractional impact radius in the lens plane, $f(x)\equiv M_{\rm disc}(\theta_I)/M_{\rm disc}(\theta_M)$  is the fraction of the embedded lens' mass projected within the impact disc of angular radius $\theta_I$, and $f_{\rm RW}(x)=1-(1-x^2)^{3/2}$ is the corresponding quantity for the removed co-moving FLRW dust sphere, see the Appendix for some examples.
At (and beyond) the boundary of the embedded lens, $f(x)=f_{\rm RW}(x)=1$. The angle
$\theta_E=\sqrt{2r_{\rm s}D_{ds}/D_dD_s}$ is the usual Einstein ring angle. Distances
$D_s$ and $D_{ds}$ are  angular diameter distances to the source measured from the observer and the deflector, respectively.
The geometrical part of the time delay $T_g$, i.e., the first term in Eq.\,(\ref{T}), has a universal form whereas the potential part  $T_p$ depends on the individual lens structure.
\footnote{ The geometrical part of the delay is the difference of arrival times of two photons starting at a fixed comoving distance $\chi_s$ and traveling \underbar{entirely} in the background cosmology. One travels on a single straight line path to the observer and the second, whose arrival time is delayed by $T_g$,  travels on two straight lines differing in direction by the deflection angle $\alpha$ at a comoving distance $\chi_d$ from the observer (see point B of Fig.\,\ref{fig:cheese}). The potential part of the time delay $T_p$ is the difference of the exiting times of two photons, red shifted to the observer, both entering the lens at point 1 at time $t_1$ but traveling on 2 separate paths to point 2. One photon travels on the two short straight lines differing in direction by the deflection angle $\alpha$ at point B. This photon travels as if it were entirely in the background cosmology and is simply reflected at point B. The delayed second photon travels on the actual null path within the lens as described by the geometry of the lens.}
To construct the Fermat potential all that is needed is a mass density profile $\rho(r,z_d)$ at lensing time, i.e., at $z_d$, for which
\be
cT_p(\theta_I,z_d)= 2(1+z_d)r_{\rm s}\int_x^1{\frac{f(z_d,x')-f_{\rm RW}(x')}{x'}{dx'}},
\label{Tp}
\ee
can be integrated.
All embedded lens properties can be constructed once the specific $T_p(\theta_I,z_d)$ is known.
For example the specific lens equation is given by a $\theta_I$-variation $\delta T(\theta_S,\theta_I)/\delta\theta_I=0$. 
This result is completely consistent with conventional lensing theory which projects the lensing mass into the lens plane; it simply accounts for the absence of lensing by the Swiss cheese void.

\section{The ISW profile from the Fermat potential}\label{sec:ISW-Fermat}
In \citep{Chen15a} we have shown that the ISW effect \cite{Sachs67,Rees68} can also be obtained by a derivative (a $z_d$-derivative) of $T$, the Fermat potential, or of $T_p$ alone since $\partial T_g/\partial z_d= 0$
\be
\frac{\Delta {\cal T}(\theta_I,z_d)}{{\cal T}}= H_d\,\frac{\partial\, T_p(\theta_I,z_d)}{\partial\, z_d}.
\label{calT}
\ee
When Eq.\,(\ref{Tp}) is inserted into  Eq.\,(\ref{calT}) two terms are separately identifiable. The first is called the time-delay part and is proportional to the potential part of the time-delay
\be
\frac{\Delta{\cal T}_{\rm T}}{{\cal T}}= \frac{H_d T_p}{1+z_d},
\label{calTT}
\ee
and a second term called the evolutionary part is present when the projected mass fraction evolves differently than the comoving background
\be
\frac{\Delta {\cal T}_{\cal E}}{{\cal T}}=(1+z_d)\frac{2r_s H_d}{c}\int_x^1\frac{dx'}{x'}\ \frac{\partial f(z_d,x^\prime)}{\partial z_d}.
\label{calTE}
\ee
In this expression $\Delta {\cal T}$ is the change in the observed CMB's temperature ${\cal T}$ caused by CMB photons passing through an evolving gravitational lens at impact angle $\theta_I$ . The cosmic-time evolution of the lens is replaced by a dependence on the redshift $z_d$ at which lensing occurs and the Hubble parameter at that redshift is denoted by $H_d=H(z_d)$.
To compute the ISW effect caused by an embedded lens, we need not only the density profile as a function of $\theta_I$  as required by  conventional lens theory \cite{Schneider92} to compute image properties, but we also need the density profile's evolution rate to compute the $z_d$-derivative.
 Because Eq.\,(\ref{calT}) contains only a first  derivative  we do not need to know the lens' history (i.e., the dynamics of its motion), only its density profile and its velocity distribution at lensing time $z_d$.

\section{The ISW profile from the lensing potential}\label{sec:ISW-Potential}
To understand the conventional expression used to compute the ISW effect for an embedded lens one starts with the spatially flat Robertson-Walker (RW) metric perturbed by a spherically symmetric lens centered at $\chi=0$. If the perturbation can be treated using the Newtonian approximation the metric can be written as
\be
ds^2=\left[1+2\Phi(t,\chi)\right](c\, dt)^2-\left[1-2\Phi(t,\chi)\right]R(t)^2\left\{(d\chi)^2+\chi^2\Bigl[(d\theta)^2+\sin^2\theta\, (d\phi)^2\Bigr]\right\},
\label{metric}
\ee
where the $\Phi(t,\chi)$ is the instantaneous Newtonian potential caused by the mass density perturbation
\be
\delta(t,\chi)\equiv \frac{\rho(t,\chi)-\overline{\rho}\,(t)}{\overline{\rho}\,(t)}
\label{delta}
\ee
i.e.,
\be
\nabla^2\Phi(t,\chi)=\frac{1}{\chi}\frac{\partial^2 }{\partial \chi^2}[\chi\,\Phi(t,\chi)]=4\pi G\, R(t)^2\overline{\rho}\,(t) \delta(t,\chi)=\frac{3}{2}\,\left(\frac{H_0}{c}\right)^2\,\Omega_m \left(\frac{\delta(t,\chi)}{R(t)}\right) .
\label{Laplace}
\ee
For an embedded lens $\delta(t,\chi)$ vanishes beyond $\chi=\chi_b$ and the boundary condition on $\Phi(t,\chi)$ is that it similarly vanish.
By writing
\be
\Phi(t,\chi)=\left(\frac{\hat{\Phi}(t,\chi)}{R(t)}\right)
\ee
equation (\ref{Laplace}) is solved by
\be
\Hat{\Phi}(t,\chi)=-\frac{3}{2}\left(\frac{H_0}{c}\right)^2\Omega_m
\left[\int^{\chi_b}_\chi \chi^\prime\left(1-\frac{\chi^\prime}{\chi}\right)\delta(t,\chi^\prime)d\chi^\prime \right].
\label{phihat}
\ee
$\Hat{\Phi}(t,\chi)$ is independent of cosmic time $t$ if the density perturbation is co-expanding with the background, i.e., if $\delta(t,\chi)$ has no dependence on $t$.

In Table 1  we have indicated the sizes of various terms that might possibly alter the form of the perturbed metric in Eq.\,(\ref{metric}) and the results below.

The conventional expression for the ISW effect is
 \bea
\frac{\Delta {\cal T}}{{\cal T}}&=& 2\,\int^{t_2}_{t_1} \frac{ \partial \Phi\bigl(t,{\chi}(t)\bigr)}{\partial t}\,dt\nonumber\\
&=&-2\,\int^{t_2}_{t_1} \frac{H(t)}{R(t)} \Hat{\Phi}\bigl(t,{\chi}(t)\bigr)\,dt+2\,\int^{t_2}_{t_1} \frac{1}{R(t)}\frac{\partial \Hat{\Phi}\bigl(t,{\chi}(t)\bigr)}{\partial t}\,dt,
\label{calTconv}
\eea
where $\chi(t)$ is the photon's comoving radial coordinate as a function of cosmic time as it passes through the lens (see Fig.\ref{fig:cheese}). For an embedded lens the integration domain is confined to the time the photon transits the Swiss cheese void. Equation\,(\ref{calTconv}) usually contains additional terms due to peculiar velocities of the emitter and/or observer as well as terms due to the emitter and/or the observer residing in local pertubations themselves. However, these terms are absent in Eq.\,(\ref{calTconv}) because we are assuming that the source and observer are comoving with the background cosmology.

\section{Equivalence}\label{sec:Equivalence}
We now show that the two terms in Eq.\,(\ref{calTconv}) are the same as the respective terms in Eqs.\,(\ref{calTT}) and (\ref{calTE}) above.
The proof amounts to showing that when performing the time integral in the conventional method the photon's path can be approximated as a straight line and that the \underbar{explicit} dependence of the integrand on the cosmic $t$ can be approximated as its value at the central time $t_d$, which corresponds to the lens' redshift $z_d$.
We write the approximate path of the photon through the comoving void as a straight line impacting the comoving lens plane at $\chi(t_d)=\chi_d\, x$,
\be
\chi(t)\approx\chi_b\sqrt{x^2+{\rm z}(t)^2}.
\label{chitz}
\ee
The actual path differs from this by terms of the order of the deflection angle $\alpha$ and because $\alpha$ is proportional to the potential $\Phi$ including such terms would be including (post-Newtonian) second order $\Phi$ terms in $\Delta{\cal T}/{\cal T}$. Such non-linear terms are small and have already been neglected in Eq.\,(\ref{metric}), see Table 1.
We next change integration variables from cosmic time to the horizontal component of fractional radial coordinate ${\rm z}$ using

\be
dt=d{\rm z}/\dot{{\rm z}}=\frac{\chi_b R(t)}{c }\left[1-2\Phi(t,\chi)\right]\,d{\rm z}\approx \frac{\chi_b R(t)}{c }\,d{\rm z},
\label{dchitz}
\ee
where the $\Phi$ term is again dropped because it would represent higher order corrections to $\Delta{\cal T}/{\cal T}$. The explicit $t$ dependence in Eq.\,(\ref{calTconv}) now becomes a function of ${\rm z}$, $t\rightarrow t({\rm z})$, via Eq.\,(\ref{dchitz}).
The first term in Eq.\,(\ref{calTconv}) thus becomes
\bea
-2\,\int^{t_2}_{t_1}  \frac{H(t)}{R(t)} \Hat{\Phi}\bigl(t,{\chi}(t)\bigr)\,dt&\approx& -2\chi_b\,\int^{+\sqrt{1-x^2}}_{-\sqrt{1-x^2}}   \frac{H\bigl(t(\rm{z})\bigr)}{c} \Hat{\Phi}\bigl(t({\rm z}),\chi_b\sqrt{x^2+{\rm z}^2}\bigr)\,d{\rm z},\nonumber\\
&\approx&-2\chi_b\, \left(\frac{H_d}{c}\right)\int^{+\sqrt{1-x^2}}_{-\sqrt{1-x^2}} \Hat{\Phi}(t_d,\chi_b\sqrt{x^2+{\rm z}^2})\,d{\rm z}.
\label{calTTconv}
\eea
The first step in Eq.\,(\ref{calTTconv}) is made using Eqs.\,(\ref{chitz}-\ref{dchitz}) and the second is made by expanding $t(\rm{z})$ about $t_d\leftrightarrow z=0$,
\be
H\bigl(t(\rm{z}) \Hat{\Phi}\bigl(t(\rm{z}),\chi\bigr)=H_d \Hat{\Phi}(t_d,\chi)+\left(H_d\frac{\partial\Hat{\Phi}(t_d,\chi)}{\partial t_d}+\dot{H}(t_d)\Hat{\Phi}(t_d,\chi)\right)\frac{\chi_b R(t_d)}{c }\,{\rm z}+{\cal O}[{\rm z}^2],
\ee
and observing that because $\chi$ is even in z and the integration range is symmetric about z$=0$, terms linear in z integrate to zero.
The ${\cal O}[{\rm z}^2]$ corrections would amount to a correction factor $\sim 10^{-4}$ for the large physical void of Table 1.

It follows by direct integration of Eq.\,(\ref{phihat}) that
\bea
\int^{+\sqrt{1-x^2}}_{-\sqrt{1-x^2}}&&\Hat{\Phi}(t_d,\chi_b\sqrt{x^2+{\rm z}^2})\,d{\rm z}=-3\left(\frac{H_0}{c}\right)^2\Omega_m \chi_b^2\nonumber\\
&\times&\int^1_xy\left[\sqrt{y^2-x^2}-y\log\left(\frac{y+\sqrt{y^2-x^2}}{x}\right)\right] \delta(t_d,\chi_b\,y)\,dy,
\label{hatPhi}
\eea
which is precisely proportional to the integrated projected mass fraction
\footnote{This result follows by computing the part of the lens mass contained within an impact cylinder of comoving radius  $y\,\chi_b$ as the mass in the central sphere of radius $y\,\chi_b$ plus the integral of the masses contained in shells of thicknesses $\delta\chi$ with radii ranging from $\chi=y\,\chi_b$ to $\chi=\chi_b$. The integral follows from observing that a shell of radius $\chi$ subtends a solid angle of $2\pi(1-\sqrt{1-y^2\chi^2_b/\chi^2})$ at the sphere's center. The two projected mass fractions $f(z_d,y)$ and $f_{RW}(y)$ are similarly obtained after dividing by the total mass $M$ and Eq.\,(\ref{ffRW}) obtains after the $y$ integration.}
\be
\int_x^1{\frac{f(t,y)-f_{\rm RW}(y)}{y}{dy}}=
3\int^1_xy\left[\sqrt{y^2-x^2}-y\log\left(\frac{y+\sqrt{y^2-x^2}}{x}\right)\right] \delta(t,\chi_b\,y)\,dy.
\label{ffRW}
\ee
When combined with Eq.\,(\ref{calTTconv}) the conclusion is that
 the first term in  Eq.\,(\ref{calTconv}) is precisely the same as $\Delta{\cal T}_{\rm T}/{\cal T}$ of Eq.\,(\ref{calTT}).

The second term in  Eq.\,(\ref{calTconv}) is approximated by a series of steps similar to those made for the first term but requires a few more steps
 \bea\label{calTEconv}
2\,\int^{t_2}_{t_1} \frac{1}{R(t)}\frac{\partial \Hat{\Phi}\bigl(t,{\chi}(t)\bigr)}{\partial t}\,dt&\approx&
2\frac{\chi_b}{c}\,\int^{+\sqrt{1-x^2}}_{-\sqrt{1-x^2}}   \frac{\partial  \Hat{\Phi}(t,\chi_b\sqrt{x^2+{\rm z}^2})}{\partial t}\,d{\rm z},\\
&\approx& 2\frac{\chi_b}{c}\, \int^{+\sqrt{1-x^2}}_{-\sqrt{1-x^2}} \frac{\partial\Hat{\Phi}(t_d,\chi_b\sqrt{x^2+{\rm z}^2})}{\partial t_d}\,d{\rm z},\nonumber\\
=-6\frac{\chi_b}{c}\left(\frac{H_0}{c}\right)^2\Omega_m \chi_b^2
&\times&\int^1_xy\left[\sqrt{y^2-x^2}-y\log\left(\frac{y+\sqrt{y^2-x^2}}{x}\right)\right] \frac{\partial\delta(t_d,\chi_b\,y)}{\partial t_d}\,dy\nonumber\\
&=&-6\left(\frac{H_0}{c}\right)^2\Omega_m \chi_b^2\int_x^1\frac{dy}{y}\ \frac{\partial f(t_d,y)}{\partial t_d},
\nonumber\\
&=&2\left(\frac{H_0}{c}\right)^2\Omega_m \chi_b^3(1+z_d)H_d\int_x^1\frac{dy}{y}\ \frac{\partial f(z_d,y)}{\partial z_d}.\nonumber
\eea
The first step in Eq.\,(\ref{calTEconv}) is made using Eqs.\,(\ref{chitz})-(\ref{dchitz}).
In the second step $t(\rm{z})$ is expanded in $z$ about $t_d$ and the integral of the term linear in $z$ vanishes. The ${\cal O}[{\rm z}^2]$ term is again too small to keep.
In the third step $\partial\Hat{\Phi}(t_d,\chi)/\partial t_d$  is related to $\partial \delta(t_d,\chi)/\partial t_d$ by differentiating  Eq.\,(\ref{hatPhi}) with respect to $t_d$. In the final step the cosmic time dependence $t_d$ is replaced by the dependence on red shift $z_d$ using
\be
\frac{d\ }{d t}=-(1+z)H(t)\frac{d\ }{d z}.
\ee
The result from Eq.\,(\ref{calTEconv}) combined with the embedding condition Eq.\,(\ref{rs}) is identical to the evolution term given in Eq.\,(\ref{calTE}).

\section{Non-compensated lenses}\label{sec:noncompensated}
We have concentrated on observational effects produced by embedded lenses to avoid the problem of keeping our models consistent with GR. We started each model  as a single lens that  exactly satisfies Einstein equations and then assumed we were dealing with lenses whose gravity field would only produce local Newtonian perturbations in the background cosmology ($\chi<\chi_b$), see Eqs.\,(\ref{metric})-(\ref{phihat}).  Such lenses are necessarily compensated by construction. The mass of a compensated lens is a contributor to the background's mean density $\overline{\rho}(t)$ and the range of its effects on passing photons is limited to its embedding radius $r_d(t)=R(t)\chi_b$.  The usual approach for lensing is to assume the lens at hand is not a contributor to the mean but an addition to it. The consequence is that the range of the lens's influence is infinite.


\section{Appendix}
Table I contains 5 mass densities $\rho(t,r)$  that can be used to  fill a comoving Swiss cheese void of  physical radius $r_b(t)=R(t)\chi_b$ in an FLRW cosmology. These densities are normalized for compensation purposes, i.e., they satisfy
\[
\int_0^{r_b}\rho(t,r)4\pi r^2\,dr=\frac{4}{3}\pi r_b^3\,\rho_{\rm RW}(t)=M.
\]
The projected mass fractions $f(t,x)$ contained in a cylinder of azimuthal radius $x * r_b $ associated with each mass density is also tabulated
\[
f(t,x)\equiv  2\pi (r_b)^3\,\int_0^x\, x^\prime\left[2\int_0^{\sqrt{1-(x^\prime)^2}}\rho(t,r_b\,\sqrt{(x^\prime )^2+{\rm z}^2})\,d{\rm z}\right]dx^\prime\Big/\frac{4}{3}\pi (r_b)^3\rho_{\rm RW}(t).
\]
By definition $f(t,x)=1$ for $x\ge 1$.

If the structure of the lens evolves differently than the background FLRW cosmology, $f(t,x)$ will depend on cosmic time. In the models that follow such a time dependence can occur if $a\equiv r_a/r_b$ is a function of $t$. We will not explicitly exhibit the $t$ dependence of $\rho$, $f$, etc., but it can be assumed there.

\begin{table}[!htbp]
\caption{Projected Mass Fractions}
\centerline{$\delta(r)$ and $\Theta(r)$ are respectively the Dirac $\delta$-function and the Heaviside step function.}
 \centerline{A time dependence occurs in $f(x)$ when the parameter $a$ depends on $t$.}
\begin{ruledtabular}
\begin{tabular}{l ||   c | c  }
lens of physical radius \ $r\le r_b$ &   $\rho(r)/\rho_{\rm RW}$  & $f(x,a),$ $\ \ x \equiv r/r_b\le 1,\ a \equiv r_a/r_b\le 1$\\ \hline\hline
Point Mass at $r=0$ & $\frac{(r_b)^3}{3}\,r^{-2}\,\delta(r)$    &1 \\  \hline
Thin Shell at $r_a\le r_b$& $\frac{(r_b)^3}{3(r_a)^2}\,\delta(r-r_a)$   & $1-\Theta(a-x)\sqrt{1-(x/a)^2}$ \\ \hline
Homogeneous Sphere $r_a\le r_b$& $(\frac{r_b}{r_a})^3\, \Theta(r_a-r)$     &  $1-\Theta(a-x)\left[\sqrt{1-(x/a)^2}\right]^3$  \\ \hline
Singular Isothermal Sphere& $\frac{(r_b)^3}{3r_a}\,r^{-2}\, \Theta(r_a-r)$    & $1-\Theta(a-x)\left\{\sqrt{1-(x/a)^2}-(x/a)\tan^{-1}\left[\frac{\sqrt{1-(x/a)^2}}{(x/a)}\right]\right\}$\\ \hline
Cubic $r_a\le r_b$ & 2$\frac{(r_b)^3}{(r_a)^6}\,r^3\, \Theta(r_a-r)$    & $1-\Theta(a-x)\Biggl\{ \sqrt{1-(x/a)^2}\left[1-\frac{3}{4}(x/a)^2\right]\left[1+\frac{1}{2}(x/a)^2\right]$\\
&&$-\frac{3}{8}(x/a)^6 \log\left[\frac{1+\sqrt{1-(x/a)^2}}{(x/a)}\right]\Biggr\} $ \\
\end{tabular}
\end{ruledtabular}
\end{table}

\eject
In Table II impact dependent integrals needed to compute  Fermat Potentials for each of the 5 lenses are tabulated
\[
{\rm FP}(x)\equiv\int^1_x\frac{f(x^\prime)}{x^\prime}dx^\prime.
\]

\begin{table}[!htbp]
\caption{Contribution to the Fermat Potential}
\begin{ruledtabular}
\begin{tabular}{l ||   c  }
lens &   ${\rm FP}(x)\equiv \int^1_x\frac{f(x^\prime)}{x^\prime}dx^\prime$ \\ \hline\hline
${\rm FP}^{{\rm PM}}(x)$   &  $-\log(x)$ \\ \hline
${\rm FP}^{{\rm TS}}(x,a)$   &  $-\log(x)+\Theta(a-x)\left\{\sqrt{1-(x/a)^2}-\log\left[\frac{1+\sqrt{1-(x/a)^2}}{(x/a)}\right]\right\}$\ \\ \hline
${\rm FP}^{{\rm HS}}(x,a)$    &  $-\log(x)+\Theta(a-x)\left\{\frac{4-(x/a)^2}{3}\sqrt{1-(x/a)^2}-\log\left[\frac{1+\sqrt{1-(x/a)^2}}{(x/a)}\right]\right\}$\\ \hline
${\rm FP}^{{\rm SIS}}(x,a)$   &  $-\log(x)+\Theta(a-x)\left\{2 \sqrt{1-(x/a)^2}-(x/a)\tan^{-1}\left[\frac{\sqrt{1-(x/a)^2}}{(x/a)}\right]-\log\left[\frac{1+\sqrt{1-(x/a)^2}}{(x/a)}\right]\right\}$\\ \hline
${\rm FP}^{{\rm Cubic}}(x,a)$ &  $-\log(x)+\Theta(a-x)\Biggl\{\frac{1}{48} \sqrt{1-(x/a)^2}\left[4-(x/a)^2\right]\left[14+3(x/a)^2\right]$ \\
 & $ -\left[1+\frac{1}{16}(x/a)^6\right] \log\left[\frac{1+\sqrt{1-(x/a)^2}}{(x/a)}\right]\Biggr\} $ \\
\end{tabular}
\end{ruledtabular}
\end{table}
\eject
From Table 1 we find
\be
f_{\rm RW}(x)=f^{\rm HS}(x,1)=1-\left(\sqrt{1-x^2}\,\right)^3
\ee
and from Table 2 we find
\be
{\rm FP_{RW}}(x)={\rm FP^{HS}}(x,1)=\frac{4-x^2}{3}\sqrt{1-x^2}-\log\left[1+\sqrt{1-x^2}\right].
\ee
In Table III  the impact dependent integrals needed to compute the potential parts of the time-delays are tabulated.

\begin{table}[!htbp]
\caption{Potential Parts of Embedded Time Delays (all satisfy ${\rm T}_p(x)\equiv 0$ for $x\ge 1$).}
\begin{ruledtabular}
\begin{tabular}{l ||   c  }
lens &   ${\rm T}_p(x)\equiv\int^1_x\frac{f(x^\prime)-f_{RW}(x^\prime)}{x^\prime}dx^\prime$ \\ \hline\hline
${\rm T}_p^{{\rm PM}}(x)$   &  $\log[\frac{1+\sqrt{1-x^2}}{x}] -\frac{4-x^2}{3}\sqrt{1-x^2}$ \\ \hline
${\rm T}_p^{{\rm TS}}(x,a)$   &  ${\rm T}_p^{{\rm PM}}(x)+\Theta(a-x)\left\{\sqrt{1-(x/a)^2}-\log\left[\frac{1+\sqrt{1-(x/a)^2}}{(x/a)}\right]\right\}$\ \\ \hline
${\rm T}_p^{{\rm HS}}(x,a)$    &   ${\rm T}_p^{{\rm PM}}(x)-{\rm T}_p^{{\rm PM}}(x/a)$\\ \hline
${\rm T}_p^{{\rm SIS}}(x,a)$   &  ${\rm T}_p^{{\rm PM}}(x)+\Theta(a-x)\left\{2 \sqrt{1-(x/a)^2}-(x/a)\tan^{-1}\left[\frac{\sqrt{1-(x/a)^2}}{(x/a)}\right]-\log\left[\frac{1+\sqrt{1-(x/a)^2}}{(x/a)}\right]\right\}$\\ \hline
${\rm T}_p^{{\rm Cubic}}(x,a)$   & $-{\rm T_p^{ PM}}(x)
+\Theta(a-x)\Biggl\{\frac{1}{48} \sqrt{1-(x/a)^2}\left[4-(x/a)^2\right]\left[14+3(x/a)^2\right]$ \\
 & $ -\left[1+\frac{1}{16}(x/a)^6\right] \log\left[\frac{1+\sqrt{1-(x/a)^2}}{(x/a)}\right]\Biggr\} $
\end{tabular}
\end{ruledtabular}
\end{table}
\centerline{\bf Superpositions}

 The potential part of the time-delay for an arbitrary superposition of normalized volume densities ($ \int_0^{r_b}\rho_i(r)4\pi r^2dr=\rho_{\rm RW}4/3\pi r^3_b=M$) is easy to compute:
\[\rho(r,z_d)=\sum_ic_i\,\rho_i(r,z_d),\]
with $\sum_ic_i=1$
\[\Rightarrow f=\sum_ic_i\,f_i\ {\rm and}\ \ {\rm T}_p=\sum_ic_i\,{\rm T}_p^i. \]


\label{lastpage}


\begin{thebibliography}{breitestes Label}

\bibitem[Sachs \& Wolfe(1967)]{Sachs67}  R. K. Sachs and A. M. Wolfe,  \apj\ {\bf 147}, 73 (1967).

\bibitem[Rees \& Sciama (1968)]{Rees68} M. J. Rees and D. W. Sciama,  Nature (London) {\bf 217} 511 (1968).

\bibitem[Granett \etal(2008a)]{Granett08}  B. R. Granett,  M. C.  Neyrinck, \& I.  Szapudi,  \apj\ Lett.  {\bf 683}, 99 (2008a).

\bibitem[Granett \etal(2008b)]{Granett08b}   B. R. Granett,  M. C.  Neyrinck, \& I.  Szapudi, arXiv.0805.2974 (2008b).

\bibitem[Planck Collaboration \etal (2014)]{Planck14}  Planck Collaboration,  P. A. R., Ade, et al., Astron. Astrophys. {\bf 571}, 19, arXiv1303.5079 (2014).

\bibitem[Inoue \& Silk(2006)]{Inoue06} K. T. Inoue \& J. Silk, \apj\ {\bf 648}, 23 (2006).

\bibitem[Rudnick \etal (2007)]{Rudnick07} L. Rudnick, S. Brown, \&  L. R. Williams, \apj\ {\bf 671}, 40 (2007).

\bibitem[Nadathur \etal(2012)]{Nadathur12} S. Nadathur, S. Hotchkiss, \&  S. Sakar, JCAP {\bf 06}, 042 (2012).

\bibitem[Hern\'andez-Monteagudo(2010)]{Hernandez10}  C. Hern\'andez-Monteagudo,  Astron. Astrophys. {\bf 520}, 101 (2010).

\bibitem[Ili\'c \etal(2013)]{Ilic13}  S. Ili\'c, M. Langer, \& M. Douspis, Astron. Astrophys. {\bf 556}, 51 (2013).

\bibitem[Cai \etal (2014)]{Cai14} Y-C. Cai, M. C. Neyrinck, I. Szapudi, S. Cole \& C. S. Frenk, \apj\ {\bf 786}, 110 (2014).

\bibitem[Chen \etal (2015b)]{Chen15b} B. Chen, R. Kantowski, and X.  Dai,   \apj\ {\bf 804}, 130 (2015).

\bibitem[Chen \& Kantowski (2015c)]{Chen15c} B. Chen \& R. Kantowski   \prd\ {\bf 91} 083014 (2015).

\bibitem[Lavaux \& Wandelt(2012)]{Lavaux12} G. Lavaux \& B. D. Wandelt,  \apj\  {\bf 754}, 109 (2012).

\bibitem[Melin \& Bartlett(2014)]{Melin14} J-B. Melin \& J. G. Bartlett, Astron. Astrophys. 578, A21 (2015). 

\bibitem[Chantavat \etal (2014)]{Chantavat14} U. Chantavat, U. Sawangwit, P. M. Sutter, \& B. D. Wandelt, arXiv.1409.3364 (2014).

\bibitem[Hamaus \etal(2014)]{Hamaus14} N. Hamaus, P. M. Sutter, G. Lavaux, \& B. D. Wandelt, JCAP {\bf 12}, 013 (2014).

\bibitem[Kantowski \etal(2015)]{Kantowski15}  R. Kantowski, B. Chen, and X. Dai,  \prd\ {\bf 91} 083004 (2015).

\bibitem[Kantowski \etal (2010)]{Kantowski10} R. Kantowski, B. Chen, and X. Dai,  \apj\  {\bf 718}, 913 (2010).

\bibitem[Chen \etal (2010)]{Chen10} B. Chen, R. Kantowski, and X.  Dai,  \prd\ {\bf 82}, 043005 (2010).

\bibitem[Chen \etal (2011)]{Chen11} B. Chen, R. Kantowski, and X.  Dai,  \prd\ {\bf 84}, 083004 (2011).

\bibitem[Kantowski \etal (2012)]{Kantowski12} R. Kantowski, B. Chen, and X. Dai, \prd\ {\bf 86}, 043009 (2012).

\bibitem[Kantowski \etal(2013)]{Kantowski13}  R. Kantowski, B. Chen, and X. Dai,  \prd\ {\bf 88}, 083001 (2013).

\bibitem[Chen \etal (2015a)]{Chen15a} B. Chen, R. Kantowski, and X.  Dai,  \apj\ {\bf 804}, 72 (2015).

\bibitem[Cooke \& Kantowski(1975)]{Cooke75} J. H. Cooke  \&  R.  Kantowski, \apj\ Lett. {\bf 195}, 11 (1975).

\bibitem[Einstein \& Straus(1945)]{Einstein45} A. Einstein and E. G. Strauss, Rev. Mod. Phys. {\bf 17}, 120 (1945).

\bibitem[Sch\"ucking(1954)]{Schucking54} E. Sch\"ucking,  Z. Phys. {\bf 137}, 595 (1954).

\bibitem[Kantowski(1969)]{Kantowski69} R. Kantowski,   \apj\ 155, {\bf 89} (1969).

\bibitem[Kottler(1918)]{Kottler18} F. Kottler, Ann. Phys. (Leipzig), 361, 401 (1918).

\bibitem[Lemaitre(1933)]{Lemaitre33} G. Lemaitre, Ann. Soc. Bruxelles {\bf A53}, 51 (1933).

\bibitem[Tolman(1934)]{Tolman34} R. C. Tolman, Proc. Natl. Acad. Sci. {\bf 20}, 169 (1934).

\bibitem[Bondi(1947)]{Bondi47} H. Bondi,  Mon. Not. R. Astron. Soc. {\bf 107}, 410 (1947).

\bibitem[Nottale(1984)]{Nottale84} L. Nottale,  Mon. Not. R. Astron. Soc.  {\bf 206}, 713 (1984).

\bibitem[Mart\'inez-Gonz\'alez \etal(1990)]{Martinez90} E. Mart\'inez-Gonz\'alez, J. L. Sanz,  \& J. Silk, \apj\ Lett. {\bf 355}, 5 (1990).

\bibitem[Panek(1992)]{Panek92} M. Panek, \apj\ {\bf 388}, 225 (1992).

\bibitem[Seljak(1996b)]{Seljak96b} U. Seljak, \apj\  {\bf 460}, 549 (1996).

\bibitem[Sakai \& Inoue(2008)]{Sakai08} N. Sakai, \&   K. T. Inoue, \prd\ {\bf 78}, 063510 (2008).

\bibitem[Valkenburg(2009)]{Valkenburg09} W. Valkenburg, JCAP {\bf 06}, 010 (2009).

\bibitem[Schneider \etal (1992)]{Schneider92} P. Schneider, J Ehlers, and E. E.  Falco,  {\it Gravitational Lenses} (Springer-Verlag, Berlin, 1992).

\bibitem[Cooray(2002)]{Cooray02}  A. Cooray,   \prd\ {\bf 65}, 083518 (2002).

\bibitem[Sch\"afer \& Bartelmann(2006)]{Schafer06} B. M. Sch\"afer \& M. Bartelmann,  Mon. Not. R. Astron. Soc.  {\bf 431}, 425 (2006).

\bibitem[Merkel \& Sch\"afer(2013)]{Merkel13} P. M. Merkel \& B. M. Sch\"afer,  Mon. Not. R. Astron. Soc.  {\bf 431}, 2433 (2013).

\bibitem[Birkinshaw \& Gull(1983)]{Birkinshaw83} M. Birkinshaw \& S. F. Gull,  Nature (London) {\bf 302}, 315 (1983).

\bibitem[Gurvits \& Mitrofanov(1986)]{Gurvits86} L. I. Gurvits \&  I. G. Mitrofanov,  Nature (London) {\bf 324}, 349 (1986).

\bibitem[Sutter \etal (2012)]{Sutter12}  P. M. Sutter, G. Lavaux, B. D.  Wandelt, \& D. H. Weinberg,  \apj\ {\bf 761}, 44 (2012).

\bibitem[Abazajian \etal(2009)]{Abazajian09}  K. N. Abazajian, et al.\  \apj\ Suppl. Ser. {\bf 182}, 543 (2009).

\bibitem[Sutter \etal (2014)]{Sutter13}  P. M. Sutter, G. Lavaux, B. D. Wandelt,   D. H. Weinberg, \&  M. S. Warren, Mon. Not. R. Astron. Soc.  {\bf 438}, 3177 (2014).

\bibitem[Nararro et al.\ (1996)]{NFW96} J. F. Navarro, C. S. Frenk,  \& S. D. White, \apj\ {\bf 462}, 563 (1996).

\bibitem[Sunyaev \& Zeldovich(1980)]{Sunyaev80}  R. A. Sunyaev,  \& Y. B. Zeldovich, Mon. Not. R. Astron. Soc.  {\bf 190}, 413 (1980).

\bibitem[Birkinshaw(1999)]{Birkinshaw99} M. Birkinshaw,  PhR\ {\bf 310}, 97 (1999).

\bibitem[Ostriker \& Cowie (1981)]{Ostriker81} J. P. Ostriker,  \& L. L. Cowie, Astrophys. J. Lett. {\bf 243}, 127 (1981).

\bibitem[Bertschinger(1985a)]{Bertschinger85a} E. Bertschinger,  \apj\ Suppl. Ser. {\bf 58}, 1 (1985a).

\bibitem[Bertschinger(1985b)]{Bertschinger85b} E. Bertschinger,  \apj\ Suppl. Ser. {\bf 58}, 39 (1985b).

\bibitem[Heath(1977)]{Heath77} D. J. Heath,  Mon. Not. R. Astron. Soc.  {\bf 179}, 351 (1977).

\bibitem[Bryan \& Norman(1998)]{Bryan98} G. L. Bryan \& M. L. Norman, \apj\ {\bf 495}, 80 (1998).

\bibitem[McBride \etal (2009)]{McBride09}  J. McBride, O. Fakhouri, \&  C.-P. Ma,  Mon. Not. R. Astron. Soc.   {\bf 398}, 1858 (2009).

\bibitem[van den Bosch(2002)]{Bosch02}  F. C. van den Bosch,  Mon. Not. R. Astron. Soc.  {\bf 331}, 98 (2002).

\bibitem[Bond \etal (1991)]{Bond91} J. R. Bond, S. Cole, G. Efstathiou, \& N. Kaiser, \apj\ {\bf 379}, 440 (1991).

\bibitem[Lacey \& Cole(1993)]{Lacey93}  C. Lacey  \&  S. Cole,  Mon. Not. R. Astron. Soc.   {\bf 262}, 627 (1993).

\bibitem[Boni \etal (2015)]{Boni15} C. De Boni, A. L. Serra, A. Diaferio, C. Giocoli, \& M. Baldi, \apj\ {\bf 818}, 188 (2016). 

\bibitem[Sheth \& van de Weygaert(2004)]{Sheth04}  R.  K. Sheth, \&  R. van de Weygaert,  Mon. Not. R. Astron. Soc.  {\bf 350}, 517 (2004).

\bibitem[Planck Collaboration \etal (2015)]{Planck15}  Planck Collaboration,  P. A. R., Ade, et al., Astron. Astrophys. {\bf 594}, A21 (2016). 

\end{thebibliography}
\end{document}